\documentclass[12pt]{article}
\usepackage{graphicx}
\def\beq{\begin{equation}}
\def\eeq{\end{equation}}
\def\bea{\begin{eqnarray}}
\def\eea{\end{eqnarray}}
\def\nn{\nonumber}
\def\nl{\nonumber\\}

\def\roughly#1{\mathrel{\raise.3ex\hbox
{$#1$\kern-.75em\lower1ex\hbox{$\sim$}}}}

\def\sla#1{\raise.15ex\hbox{$/$}\kern-.57em #1}

\def\bd{B_d^0}
\def\bs{B_s^0}
\def\bsbar{{\bar B}_s^0}
\def\btod{{\bar b} \to {\bar d}}
\def\btos{{\bar b} \to {\bar s}}
\def\fT{f_T}
\def\fL{f_L}
\def\fTfL{f_T/f_L}
\def\BKstar{B\to\phi K^*}
\def\bsss{{\bar b} \to {\bar s} s {\bar s}}

\begin{document}

\begin{flushright}  
UMISS-HEP-2011-01\\
UdeM-GPP-TH-11-197 \\
\end{flushright}

\begin{center}
\bigskip
{\Large \bf \boldmath Searching for New Physics with $B$-Decay \\ Fake Triple Products } \\

\bigskip
{\large Alakabha Datta $^{a,}$\footnote{datta@phy.olemiss.edu},
Murugeswaran Duraisamy $^{a,}$\footnote{duraism@phy.olemiss.edu} \\
and David London $^{b,}$\footnote{london@lps.umontreal.ca}}
\\
\end{center}

\begin{flushleft}
~~~~~~$a$: {\it Department of Physics and Astronomy, 108 Lewis Hall, }\\ 
~~~~~~~~~~{\it University of Mississippi, Oxford, MS 38677-1848, USA}\\
~~~~~~$b$: {\it Physique des Particules, Universit\'e de
Montr\'eal,}\\
~~~~~~~~~~{\it C.P. 6128, succ. centre-ville,
Montr\'eal, QC, Canada H3C 3J7}
\end{flushleft}
\begin{center} 
\vskip0.5cm
{\Large Abstract\\}
\vskip3truemm

\parbox[t]{\textwidth} {In pure-penguin $\btos$ $B\to V_1 V_2$ decays
  ($V_{1,2}$ are vector mesons), $\fTfL \simeq 1$ has been observed
  ($\fT$ ($\fL$) is the polarization fraction of transverse
  (longitudinal) decays). Explanations of this unexpectedly large
  result have been given within the standard model (SM) and with new
  physics (NP). In this paper, we show that these two explanations can
  be partially distinguished through the triple products (TP's) in
  these transitions. In particular, the SM predicts one of the two
  fake, CP-conserving TP's to be small ($|A_T^{(2)}| \le 9\%$), while
  NP often gives larger values for $|A_T^{(2)}|$. We discuss the
  implications of the measurements of both fake TP's in $\BKstar$ --
  the present data prefer a SM explanation of $\fTfL$ -- and provide
  the SM predictions for $\bs\to\phi\phi$.}
\end{center}

\thispagestyle{empty}
\newpage
\setcounter{page}{1}
\baselineskip=14pt

Over the past 10-15 years, a great deal of effort has been put into
measuring CP violation in the $B$ system. The great majority of these
measurements have been of direct and indirect CP asymmetries in $B$
decays. As always, the goal is to find a discrepancy with the
predictions of the standard model (SM). To date, the measurements are
generally in agreement with the SM. However, there are some small
hints of disagreements in some $\btos$ decays.

Some time ago, it was pointed out that there is another signal of CP
violation in $B\to V_1 V_2$ ($V_{1,2}$ are vector mesons) -- a triple
product (TP) \cite{Valencia,TPDL}.  In the rest frame of the $B$, the
TP takes the form ${\vec q} \cdot ({\vec\varepsilon}_1 \times
{\vec\varepsilon}_2)$, where ${\vec q}$ is the difference of the
momenta of the final vector mesons, and ${\vec\varepsilon}_1$ and
${\vec\varepsilon}_2$ are the polarizations of $V_1$ and $V_2$.  The
TP is odd under both parity and time reversal, and thus constitutes a
potential signal of CP violation.

The most general Lorentz-covariant amplitude for the decay $B(p) \to
V_1(k_1,\varepsilon_1) + V_2(k_2,\varepsilon_2)$ is given by
\cite{Valencia,TPDL}
\beq
M = a \, \varepsilon_1^* \cdot \varepsilon_2^* + {b \over m_B^2}
(p\cdot \varepsilon_1^*) (p\cdot \varepsilon_2^*) + i {c \over m_B^2}
\epsilon_{\mu\nu\rho\sigma} p^\mu q^\nu \varepsilon_1^{*\rho}
\varepsilon_2^{*\sigma} ~,
\label{amp1}
\eeq
where $q\equiv k_1 - k_2$. The quantities $a$, $b$ and $c$ are complex
and will in general contain both CP-conserving strong phases and
CP-violating weak phases.  In $B\to V_1 V_2$ decays, the final state
can have total spin 0, 1 or 2, which correspond to the $V_1$ and $V_2$
having relative orbital angular momentum $l=0$ ($s$ wave), $l=1$ ($p$
wave), or $l=2$ ($d$ wave), respectively.  The $a$ and $b$ terms
correspond to combinations of the parity-even $s$- and $d$-wave
amplitudes, while the $c$ term corresponds to the parity-odd $p$-wave
amplitude.

In order to obtain experimental information about the TP, one uses the
linear polarization basis.  Here, one decomposes the decay amplitude
into components in which the polarizations of the final-state vector
mesons are either longitudinal ($A_0$), or transverse to their
directions of motion and parallel ($A_\|$) or perpendicular
($A_\perp$) to one another. $A_0$, $A_\|$, $A_\perp$ are related to
$a$, $b$ and $c$ of Eq.~(\ref{amp1}) via \cite{TPDL}
\beq
A_\| = \sqrt{2} a ~,~~~ A_0 = -a x - {m_1 m_2 \over m_B^2} b
(x^2 - 1) ~,~~~ A_\perp = 2\sqrt{2} \, {m_1 m_2 \over m_B^2} c 
\sqrt{x^2 - 1} ~,
\label{Aidefs}
\eeq
where $x = k_1 \cdot k_2 / (m_1 m_2)$ ($m_1$ and $m_2$ are the masses
of $V_1$ and $V_2$, respectively.). Now, in the rest frame of the $B$,
the $c$ term of Eq.~(\ref{amp1}) is proportional to the TP ${\vec q}
\cdot ({\vec\varepsilon}_1 \times {\vec\varepsilon}_2)$. Thus, there
are two TP terms in $|M|^2$, proportional to ${\rm Im}(c a^*)$ and
${\rm Im}(b a^*)$ \cite{TPDL}. Equivalently, from the above equation,
the two TP's are proportional to linear combinations of ${\rm
  Im}(A_\perp A_0^*)$ and ${\rm Im}(A_\perp A_\|^*)$. (Note that this
is to be expected -- the TP is parity-odd, and one can generate such
an effect through the interference of either the $l$-even $s$-wave or
$d$-wave state (i.e.\ $A_0$ or $A_\|$) with the $l$-odd $p$-wave state
($A_\perp$).)

Assuming that $V_{1,2}$ both decay into pseudoscalars, i.e.\ $V_1 \to
P_1 P_1'$, $V_2 \to P_2 P_2'$, the angular distribution of $B\to V_1
V_2$ is then given by \cite{DDLR,CW}
\bea
{d\Gamma \over d\cos\theta_1 d\cos\theta_2 d\phi} & = & N \left(
|A_0|^2 \cos^2\theta_1 \cos^2\theta_2 + {|A_\perp|^2 \over 2}
\sin^2\theta_1 \sin^2\theta_2 \sin^2 \phi \right. \nn\\
& & \hskip-1.0truein 
+ {|A_{\|}|^2 \over 2} \sin^2\theta_1 \sin^2\theta_2 \cos^2
\phi + {{\rm Re}(A_0 A_\|^*) \over 2\sqrt{2}} \sin 2\theta_1 \sin
2\theta_2 \cos\phi \nn\\
& &  \hskip-1.0truein \left. 
- {{\rm Im}(A_\perp A_0^*) \over 2\sqrt{2}} \sin 2\theta_1
\sin 2\theta_2 \sin\phi - {{\rm Im}(A_\perp A_\|^*) \over 2}
\sin^2\theta_1 \sin^2\theta_2 \sin 2\phi \right),
\label{angdist}
\eea
where $\theta_1$ ($\theta_2$) is the angle between the directions of
motion of the $P_1$ ($P_2$) in the $V_1$ ($V_2$) rest frame and the
$V_1$ ($V_2$) in the $B$ rest frame, and $\phi$ is the angle between
the normals to the planes defined by $P_1 P_1'$ and $P_2 P_2'$ in the
$B$ rest frame. (For other decays of the $V_1$ and $V_2$ (e.g.\ into
$e^+ e^-$, $P \gamma$ or three pseudoscalars), one will obtain a
different angular distribution, see Refs.~\cite{DDLR,CW,KP}.)
The key point is that, by performing a full angular analysis of $B\to
V_1 V_2$, one can obtain ${\rm Im}(A_\perp A_0^*)$ and ${\rm
  Im}(A_\perp A_\|^*)$, i.e.\ both TP's, from Eq.~(\ref{angdist}) above.

Now, above we indicated that TP's are a signal of CP violation. This
is not quite accurate. As already noted, in general the $A_i$
($i=0,\|,\perp$) possess both weak (CP-odd) and strong (CP-even)
phases. Thus, ${\rm Im}(A_\perp A_0^*)$ and ${\rm Im}(A_\perp A_\|^*)$
can both be nonzero even if the weak phases vanish. In order to obtain
a true signal of CP violation, one has to compare the $B$ and ${\bar
  B}$ decays. The amplitude for ${\bar B}(p) \to {\bar
  V}_1(k_1,\varepsilon_1) + {\bar V}_2(k_2,\varepsilon_2)$ can be
obtained by operating on Eq.~(\ref{amp1}) with CP. This yields
\beq
{\bar M} = {\bar a} \, \varepsilon_1^* \cdot \varepsilon_2^* + {{\bar
    b} \over m_B^2} (p\cdot \varepsilon_1^*) (p\cdot \varepsilon_2^*)
- i {{\bar c} \over m_B^2} \epsilon_{\mu\nu\rho\sigma} p^\mu q^\nu
\varepsilon_1^{*\rho} \varepsilon_2^{*\sigma} ~,
\eeq
in which ${\bar a}$, ${\bar b}$ and ${\bar c}$ are equal to $a$, $b$
and $c$, respectively, except that the weak phases are of opposite
sign. Thus, the above equation can be obtained from Eq.~(\ref{amp1})
by changing $a \to {\bar a}$, $b \to {\bar b}$ and $c \to -{\bar c}$.
Similarly, the angular distribution of this decay is the same as that
in Eq.~(\ref{angdist}), with $A_0 \to {\bar A}_0$, $A_\| \to {\bar
  A}_\|$ and $A_\perp \to -{\bar A}_\perp$\footnote{Another way to see
  this is to note that when one acts on the $B$ decay amplitude
  [Eq.~(\ref{amp1})] with CP, there is a factor $(-1)^l$ for each
  term, where $l$ is the relative angular momentum of the two vector
  mesons. Since the $c$-term corresponds to a state with $l=1$, the
  ${\bar c}$ term, which is related to ${\bar A}_\perp$, has an
  additional factor of $-1$ associated with it. As a consequence, the
  TP's in the angular distribution, which are proportional to ${\bar
    A}_\perp$, also have a factor of $-1$.}, in which the ${\bar A}_i$
are obtained from the $A_i$ by changing the sign of the weak phases.

The point is that the TP's for the ${\bar B}$ decay are $-{\rm
  Im}({\bar A}_\perp {\bar A}_0^*)$ and $-{\rm Im}({\bar A}_\perp
{\bar A}_\|^*)$. The true (CP-violating) TP's are then given by
$\frac12[{\rm Im}(A_\perp A_0^*) - {\rm Im}({\bar A}_\perp {\bar
    A}_0^*)]$ and $\frac12[{\rm Im}(A_\perp A_\|^*) - {\rm Im}({\bar
    A}_\perp {\bar A}_\|^*)]$. But there are also fake (CP-conserving)
TP's, due only to strong phases of the the $A_i$'s. These are given by
$\frac12[{\rm Im}(A_\perp A_0^*) + {\rm Im}({\bar A}_\perp {\bar
    A}_0^*)]$ and $\frac12[{\rm Im}(A_\perp A_\|^*) + {\rm Im}({\bar
    A}_\perp {\bar A}_\|^*)]$. For the fake TP's, it is necessary to
distinguish $B$ and ${\bar B}$, i.e.\ untagged samples contain no fake
TP's \cite{DunFlei}.

In order to illustrate characteristics of the true and fake TP's,
suppose that there are two amplitudes ${\cal A}_1$ and ${\cal A}_2$
contributing to a given decay, and that the TP is proportional to
${\rm Im}({\cal A}_1 {\cal A}_2^*$). It is straightforward to show that
\bea 
TP_{true} \propto \sin{\phi}\cos{\delta} ~, \nl
TP_{fake} \propto \cos{\phi}\sin{\delta} ~,
\label{TPangles}
\eea
where $\phi$ and $\delta$ are, respectively, the relative weak and
strong phases between ${\cal A}_1$ and ${\cal A}_2$. As is clear from
these expressions, the true TP requires a nonzero $\phi$ and is
relatively insensitive to $\delta$. That is, as with any genuine
CP-violating effect, the interference of two amplitudes with a
relative weak phase is required. On the other hand, the fake TP
requires only a nonzero strong-phase difference $\delta$, and can be
nonzero even if the weak-phase difference $\phi$ vanishes. Since the
linear polarization amplitudes will, in general, have different strong
phases, this will lead to nonzero fake TP's for all decays.

For the two TP's of Eq.~(\ref{angdist}), we define
\beq
A_T^{(1)} \equiv \frac{{\rm Im}(A_\perp
  A_0^*)}{A_0^2+A_\|^2+A_\perp^2} ~~,~~~~
A_T^{(2)} \equiv \frac{{\rm Im}(A_\perp
A_\|^*)}{A_0^2+A_\|^2+A_\perp^2} ~.
\label{TPmeasure}
\eeq 
The corresponding quantities for the charge-conjugate process,
$\bar{A}_T^{(1)}$ and $\bar{A}_T^{(2)}$, are defined similarly, but
with a multiplicative minus sign.  Consider now $B \to V_1 V_2$ decays
in which the final vector mesons are light: $m_{1,2} \ll m_B$. In
Ref.~\cite{TPDL} it was shown that, in the SM within factorization,
\beq
{|A_{\|,\perp}| \over |A_0|} = O\left( \frac{m_{1,2}}{m_B} \right) ~.
\eeq
That is, the transverse amplitudes are naively expected to be much
smaller than the longitudinal amplitude. This implies that, in
general, $|A_T^{(2)}| \ll |A_T^{(1)}|$.

One also expects that, in $B\to V_1 V_2$, the fraction of transverse
decays, $\fT$, is much less than the fraction of longitudinal decays,
$\fL$. However, it was observed that these two fractions are roughly
equal in the decay $\BKstar$: $\fTfL \simeq 1$ \cite{phiK*}. There are
two possible explanations of this. The first is that the SM is still
valid, but one must go beyond the minimal version. One scenario is
that nonfactorizable QCD-factorization penguin-annihilation effects
are important \cite{Kagan}. A second scenario involves nonperturbative
rescattering \cite{soni, SCET}.  Alternatively, one can explain the
$\fTfL$ measurement by introducing physics beyond the SM. Suppose
there is a new-physics (NP) contribution to the $\bsss$ quark-level
amplitude. If the NP operator has the structure $(1-\gamma_5) \otimes
(1-\gamma_5)$ or $\sigma (1-\gamma_5) \otimes \sigma (1-\gamma_5)$
(denoted $ST_{LL}$ below), or $(1+\gamma_5) \otimes (1+\gamma_5)$ or
$\sigma (1+\gamma_5) \otimes \sigma (1+\gamma_5)$ ($ST_{RR}$), this
will contribute dominantly to $\fT$ in $\BKstar$ and not to $\fL$
\cite{BphiK*NP,adpol}. One can therefore reproduce the measured value
of $\fTfL$ if the NP amplitude has the right size. In this paper, we
do not assess the advantages and disadvantages of the two
explanations. Rather, our aim is to propose a way of distinguishing
them.

$B\to V_1 V_2$ decays can be separated into four types. These include
$\btos$ transitions: (i) pure penguin (e.g.\ $\BKstar$), (ii)
tree and penguin contributions (e.g.\ $B\to\rho K^*$), and $\btod$
transitions: (i) pure penguin (e.g.\ $B\to K^* {\bar K}^*$), (ii) tree
and penguin contributions (e.g.\ $B\to\rho \rho$). The polarizations
have been measured for the decays in parentheses (and others
\cite{hfag}). The results are shown in Table 1. 

\begin{table}
\null~~~~~~~~~~~~~~~~~~~~~~~
\begin{tabular}{|c|c|c|}
\hline
Decay & Final State & $\fL$  \\
\hline
$\BKstar$ \cite{phiK*} & $\phi K^{*0}$ &  $0.480 \pm 0.030$ \\
& $\phi K^{*+}$ & $0.50 \pm 0.05$ \\
\hline
$B\to\rho K^*$ \cite{rhoK*} & $\rho^0 K^{*0}$ & $0.57 \pm 0.12$ \\
& $\rho^+ K^{*0}$ & $0.48 \pm 0.08$ \\
\hline
$B\to K^* {\bar K}^*$ \cite{K*Kbar*} & $K^{*0} {\bar K}^{*0}$ & $0.80^{+0.12}_{-0.13}$ \\
& $K^{*+} {\bar K}^{*0}$ & $0.75^{+0.16}_{-0.26}$ \\
\hline
$B\to\rho \rho$ \cite{rhorho} & $\rho^+ \rho^-$ & $0.978^{+0.025}_{-0.022}$ \\
& $\rho^0 \rho^0$ & $0.75^{+0.12}_{-0.15}$ \\
& $\rho^+ \rho^0$ & $0.950 \pm 0.016$ \\
\hline
\end{tabular}
\caption{Longitudinal polarization fraction $\fL$ for various $B\to
  V_1 V_2$ decays, taken from Ref.~\cite{hfag}.
\label{table:parameters}}
\end{table}

As noted above, there is an effect in the $\btos$ penguin amplitude
which leads to $\fTfL \simeq 1$. There is a similar, though weaker,
effect in the $\btod$ penguin amplitude giving $\fTfL \simeq 1/3$. The
data suggest that the tree amplitude(s) reproduce the naive
expectations, i.e.\ the transverse amplitudes are much smaller than
the longitudinal amplitude. Thus, in $\bd\to\rho^+ \rho^-$ and
$B^+\to\rho^+ \rho^0$, which are $\btod$ decays with both tree and
penguin contributions, we have $\fTfL \simeq 0$. This is because the
color-allowed tree amplitude is the dominant contribution. (In
$\bd\to\rho^0 \rho^0$, the color-suppressed tree amplitude is smaller,
and the contribution of the $\btod$ penguin amplitude leads to $\fTfL
\simeq 1/3$.) And in the $\btos$ decay with both tree and penguin
contributions, the tree amplitude, though nonzero, is
subdominant. This gives a value for $\fTfL$ which is slightly smaller
than that for the pure penguin $\btos$ decay. The upshot of all of
this is that there are three classes of decays in which the transverse
polarizations are reasonably large. Therefore, for these decays, we
have $|A_T^{(2)}| \simeq |A_T^{(1)}|$, contrary to our naive expectation.

However, there is more, and this is the main point of this paper. It
is also possible to express the polarization amplitudes using the
helicity formalism. Here, the transverse amplitudes are written as
\bea
A_\| &=& \frac{1}{\sqrt{2}} (A_+ + A_-) ~, \nn\\
A_\perp &=& \frac{1}{\sqrt{2}} (A_+ - A_-) ~.
\eea
The key observation is the following. Due to the fact that the weak
interactions are left-handed, i.e.\ the couplings are $V-A$, the
helicity amplitudes obey the hierarchy \cite{Kagan,BRY}
\beq
\left\vert \frac{A_+}{A_-} \right\vert = \frac{\Lambda_{QCD}}{m_b} ~.
\eeq
Thus, in the heavy-quark limit, $A_+$ is negligible compared to $A_-$,
so that $A_\| = -A_\perp$. But in this case, $A_T^{(2)}$, which is
proportional to ${\rm Im}(A_\perp A_\|^*)$, vanishes. This means that
if the large $\fTfL$ observed in several $B\to V_1 V_2$ decays is due
to the SM, $A_T^{(2)} = 0$ should be found. On the other hand, suppose
that the large $\fTfL$ is due to NP. If the new interactions have a
different weak phase from the SM, they can be detected using the true
TP's (of $A_T^{(1)}$ or $A_T^{(2)}$). However, the NP could have the
same weak phase as the SM, so that the true TP's vanish
[Eq.~(\ref{TPangles})]. It may therefore not be ideal to concentrate
on the true TP's -- it also may be useful to measure the fake TP
constructed from $A_T^{(2)}$ and $\bar{A}_T^{(2)}$. As we will see
below, it is possible to partially distinguish the SM from NP through
the measurement of the fake $A_T^{(2)}$ TP.

Of course, there are corrections to the prediction that $A_T^{(2)} =
0$, since the heavy-quark limit is just an approximation. Below, we
take these corrections into account, and estimate $A_T^{(2)} $ for the
pure-penguin $\btos$ decays $\BKstar$ and $\bs\to\phi\phi$. We
also comment on the size of $A_T^{(2)} $ for other $\btos$ and $\btod$
transitions.

We take $A_\lambda = |A_\lambda| e^{i \delta_\lambda}$ ($\lambda = 0,
\pm $), and define $r_T \equiv |A_+/A_-|$. $A_T^{(2)}$ is then given
by
\beq
A_T^{(2)} = \frac{ r_T  f_T}{ (1 + r^2_T )} \sin{(\delta_+-\delta_-)} ~,
\label{AT2eqn}
\eeq
where the polarization fractions are
\beq
f_i = \frac{|A_i|^2}{|A_0|^2+|A_\perp|^2+|A_\parallel|^2} ~, ~~~~ i=0,\perp, \parallel ~,
\eeq
with $f_T = f_\perp + f_\parallel$. In $\btos$ transitions, all
contributions to the decay are proportional to the
Cabibbo-Kobayashi-Maskawa (CKM) factors $V_{tb}^*V_{ts}$,
$V_{cb}^*V_{cs}$, or $V_{ub}^*V_{us}$. The term $V_{cb}^*V_{cs}$ can
be eliminated in terms of the other two using the unitarity of the CKM
matrix. Furthermore, although $V_{ub}^*V_{us}$ has a large weak phase,
its magnitude is greatly suppressed relative to that of
$V_{tb}^*V_{ts}$. In pure-penguin $\btos$ decays, the $V_{ub}^*V_{us}$
term is negligible, to a good approximation. That is, there is
effectively only one weak amplitude (i.e.\ $(\delta_+-\delta_-)$ is
purely a strong phase), and so all CP-violating effects are
tiny. Thus, $A_T^{(2)}=-\bar{A}_T^{(2)} $ and so $A_T^{(2)} $ is by
itself a fake TP.

In order to estimate the size of $A_T^{(2)} $, we proceed as
follows. First, within QCD factorization \cite{Beneke:1999br}, $r_T$
is expected to be about 4\%. When the penguin-annihilation amplitude
is added, $r_T$ is increased to lie in the range 5\%-15\%. Second, it
is straightforward to show that
\bea
\frac{[(1-r^2_T)^2+ 4 r^2_T
    \sin^2{(\delta_+-\delta_-)}]^{1/2}}{1+r^2_T+ 2 r_T
  \cos{(\delta_+-\delta_-)}} &=& \sqrt{\frac{f_\perp}{f_\parallel}} ~.
\label{fperp_par}
\eea
Given the experimental values for ${f_\perp}$ and ${f_\parallel}$, the
above equation provides a constraint on $r_T$ and the phase
$(\delta_+-\delta_-)$.

\begin{figure}
\includegraphics[width=0.49\linewidth]{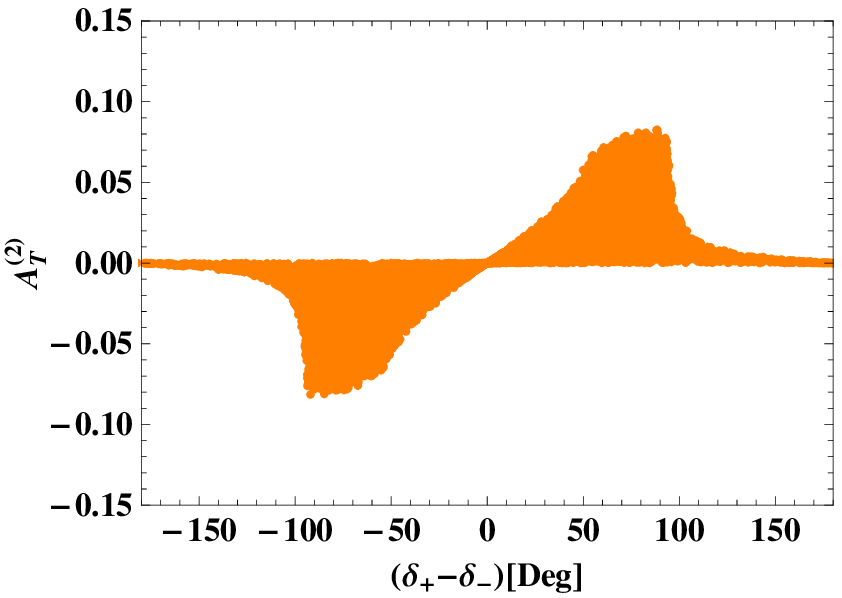}
\includegraphics[width=0.49\linewidth]{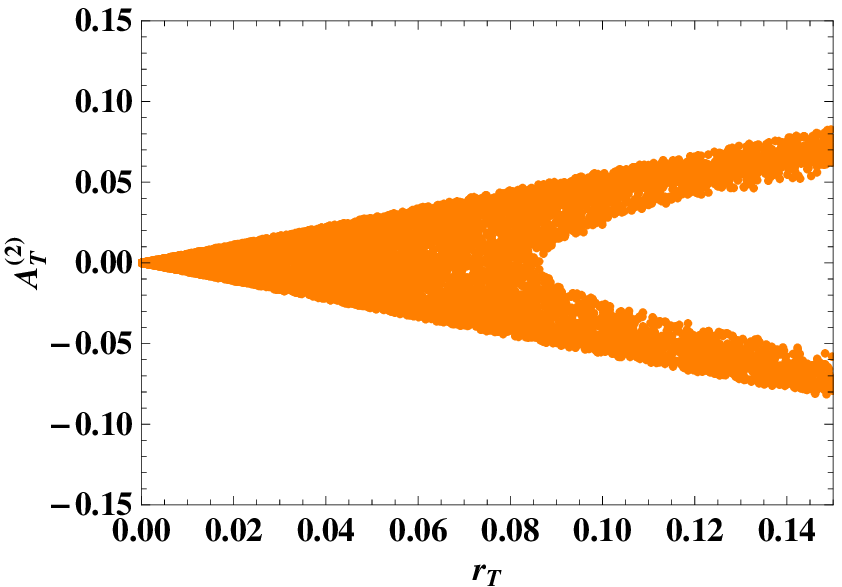}
\caption{The left (right) panel of the figure shows $A_T^{(2)}$ for
  the decay $B_d \to \phi K^{*0}$ as a function of
  $(\delta_+-\delta_-)$ ($r_T$).}
\label{fig:TPphikstAT2}
\end{figure}

We begin with $\BKstar$. If desired, one can avoid tagging altogether
by considering charged-$B$ decays, or by using self-tagging decays of
the $K^{*0}$ in $\bd\to\phi K^{*0}$. The estimate for $A_T^{(2)}$ is
found using Eq.~(\ref{AT2eqn}).  $r_T$ is varied in the range (0.05,
0.15), and the phases $\delta_\pm$ in the range (0, $2\pi$). The
constraint of Eq.~(\ref{fperp_par}) is imposed using the measured
polarization fractions $f_\perp = 0.241 \pm 0.029$ and $f_\parallel =
1 - f_L - f_\perp = 0.279 \pm .042$ \cite{hfag}. The experimental
numbers are varied within their $\pm 1 \sigma$ errors. The result is
shown in Fig.~\ref{fig:TPphikstAT2}. There we see that $|A_T^{(2)}|
\le 9\%$ is predicted.

This prediction can be compared with the experimental result.
$A_T^{(2)}$ has not been explicitly measured, but its value can be
deduced using other measurements. The relevant $B_d \to \phi
K^{*0}$ polarization observables are shown in Table
\ref{Kstphipoldata}. Here, the relative phases between
$A_{\perp,\parallel}$ and $A_0$, denoted $\phi_\perp$ and
$\phi_\parallel$, are defined to be
\bea
\label{poleq}
 \phi_i &=& arg \frac{A_i}{A_0}-\pi~ \mathrm{sign} ( arg \frac{A_i}{A_0}) ~, ~~~~ i=\perp, \parallel ~.
\eea
We follow the convention of Ref.~\cite{babar04} for the polarization
fractions, and of Ref.~\cite{BRY} for the phases, defining
\bea
f^Q_L &\equiv& f_L (1+ Q~ A^{0}_{CP}) ~,~~f^Q_\perp \equiv f_\perp (1+ Q~ A^{\perp}_{CP}) ~,\nl
\phi^Q_h &=& \phi_h + Q ~\Delta \phi_h ~, ~~~~ h = \parallel, \perp ~.
\eea
Here, $Q=1$ ($-1$) for $\bar{B}^0$ ($B^0$). Using the numbers above we
can calculate $A_T^{(2)}$:
\bea
A_T^{(2)} &=& \frac{1}{2}(A_{T,B}^{(2)} - \bar{A}_{T,{\bar{B}}}^{(2)} )= 0.002\pm 0.049 ~.
\eea

The measured value of $A_T^{(2)}$ is therefore in agreement with the
SM prediction. Indeed, it is consistent with zero. What does this say
about the NP explanations of the large observed value of $\fTfL$? In
the heavy-quark limit, $A_+ = 0$ in the $ST_{LL}$ scenario, so that
$A_\| = -A_\perp$ (as in the SM) and $A_T^{(2)} = 0$. Similarly,
$ST_{RR}$ predicts that $A_- = 0$, so that $A_\| = A_\perp$ and
$A_T^{(2)} = 0$. Thus, the result that $A_T^{(2)} \simeq 0$ is
consistent with $ST_{LL}$. It also appears to be consistent with
$ST_{RR}$. However, the SM and $ST_{RR}$ make very different
predictions for $A_+$ and $A_-$. Since both the SM and $ST_{RR}$
operators are present in this NP scenario -- the value of $f_L$
confirms the importance of the SM contribution -- the predicted value
of $A_T^{(2)}$ is nonzero. Thus, the measurement of $A_T^{(2)} \simeq
0$ rules out $ST_{RR}$, or at least strongly constrains
it. Furthermore, in real model calculations (e.g.\ in the
two-Higgs-doublet model \cite{DIL}), in general both $ST_{LL}$ and
$ST_{RR}$ operators appear, so that neither $A_+$ nor $A_-$ is zero,
and $A_T^{(2)} \ne 0$. As above, such NP scenarios are generally ruled
out. (Note that even if the NP operators have new weak phases, this
will not significantly affect the fake $A_T^{(2)}$ TP, see
Eq.~(\ref{TPangles}). The one exception is if the new phase is near
$\frac{\pi}{2}$ or $\frac{3\pi}{2}$. In this case, the fake TP is
small, and the NP must be detected through a true TP, which is
maximal.) We therefore see that the measurement of the fake
$A_T^{(2)}$ TP allows us to partially differentiate the SM from the NP
explanations of $\fTfL$. The present $\BKstar$ data suggest that the SM
is preferred over NP.

\begin{table}
\centering
\begin{tabular}{|l|l|}\hline
\multicolumn{2}{|c|}{Polarization fractions}\\ \hline
 $f_L = 0.480 \pm 0.030$ &  $f_\perp = 0.241 \pm 0.029$  \\ \hline
\multicolumn{2}{|c|}{Phases}\\ \hline
$\phi_\parallel(rad) = 2.40^{+0.14}_{-0.13}$ & $\phi_\perp(rad) =2.39 \pm 0.13$ \\ \hline
$\Delta \phi_\parallel(rad) = 0.11 \pm 0.13$ & $\Delta \phi_\perp(rad) = 0.08 \pm 0.13$ \\ \hline
\multicolumn{2}{|c|}{CP asymmetries}\\ \hline
$A^0_{CP} = 0.04 \pm 0.06$ & $A^\perp_{CP} = -0.11 \pm 0.12$ \\ \hline
\end{tabular}
\caption{$B_d \to \phi K^{*0}$ polarization observables \cite{hfag}.}
\label{Kstphipoldata}
\end{table}

\begin{figure}
\includegraphics[width=0.49\linewidth]{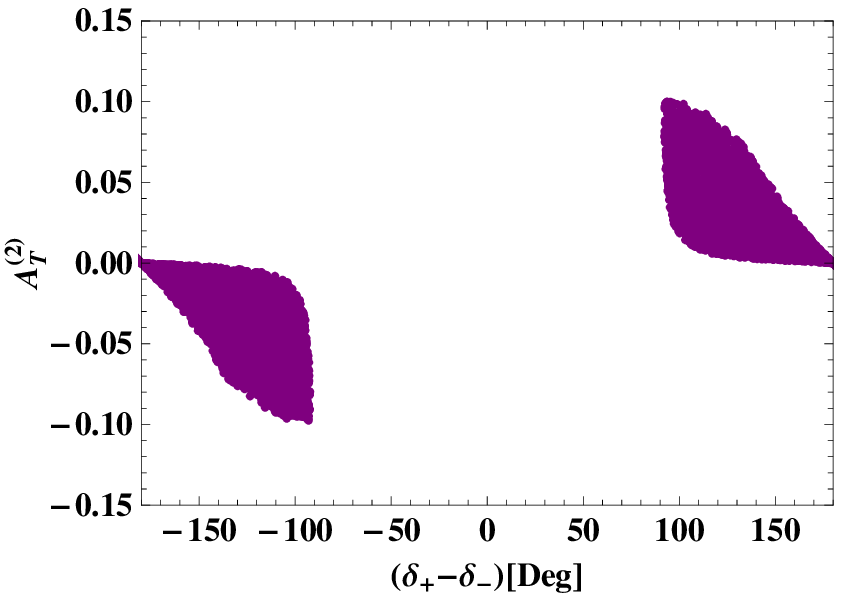}
\includegraphics[width=0.49\linewidth]{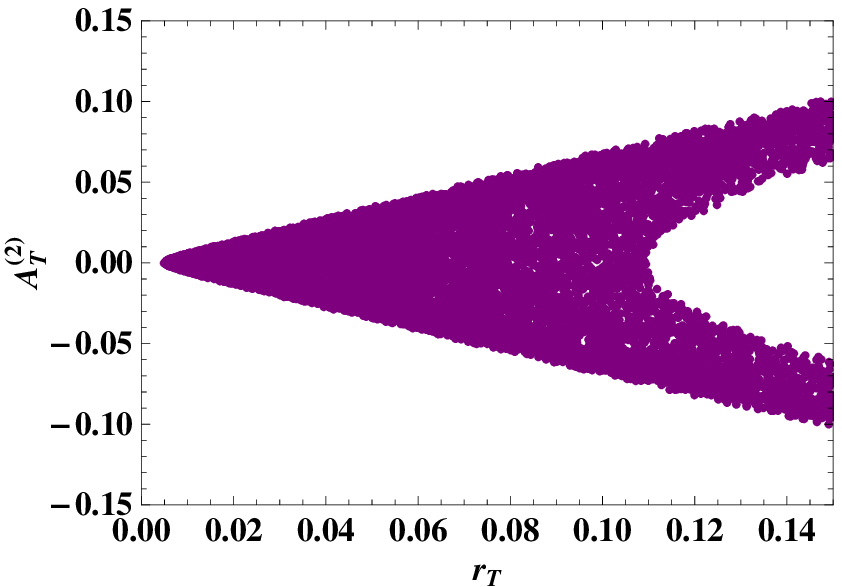}
\caption{The left (right) panel of the figure shows $A_T^{(2)}$ for
  the decay $B_s \to \phi \phi$ as a function of
  $(\delta_+-\delta_-)$ ($r_T$).
  \label{fig:TPphiphiAT2}}
\end{figure}

We now turn to $B_s \to \phi \phi$. In this case, tagging is necessary
to distinguish the $\bs$ and $\bsbar$ decays. Furthermore,
$\bs$-$\bsbar$ mixing must be taken into account. Within the SM, in
which the weak phase of the mixing is negligible, the TP terms of
Eq.~(\ref{angdist}) are modified as follows \cite{Dunietz}:
\beq
{\rm Im}(A_\perp A_{0,\|}^*) \to e^{-\Gamma t} \left( {\rm Im}(A_\perp A_{0,\|}^*) \cos \Delta m t 
- {\rm Re}(A_\perp A_{0,\|}^*) \sin \Delta m t \right) ~,
\eeq
where we have set the mixing phase to 0.  As before, we use
Eq.~(\ref{AT2eqn}) to estimate $A_T^{(2)}$, taking $r_T$ and
$\delta_\pm$ in the ranges (0.05, 0.15) and (0, $2\pi$),
respectively. The CDF data for the polarization observables for this
decay are \cite{Rescigno:2011qr}
\bea
\label{cdfpol}
f_L &=& 0.348 \pm 0.041 (stat)\pm 0.021 (syst) ~, \nl
   f_\parallel &=& 0.287 \pm 0.043 (stat)\pm 0.011 (syst) ~, \nl
   f_\perp &=& 0.365 \pm 0.044 (stat)\pm 0.027 (syst) ~, \nl
f_T &=&   0.652 \pm 0.041 (stat)\pm 0.021 (syst) ~.
\eea
These are used to impose the constraint of Eq.~(\ref{fperp_par}) (the
experimental numbers are varied within $\pm 1 \sigma$). The result is
shown in Fig.~\ref{fig:TPphiphiAT2}. The prediction is that
$|A_T^{(2)}| \le 10\%$.

It is interesting to compare the decays $\BKstar$ and $B_s \to
\phi \phi$. These are the same in the flavor SU(3) limit. However, at
present there are signifant differences. For example, the polarization
fraction $f_L$ differs between the two decays by more than $3\sigma$.
(Still, SU(3) breaking, which can have an effect of $\sim 25\%$, could
account for this.)  Due to the fact that $f_L$ is relatively small in
$B_s \to \phi \phi$, a sizeable fraction of the
$(\delta_{+}-\delta_{-})$ space is not allowed in
Fig.~\ref{fig:TPphiphiAT2}. If we ignore the differences between the
two decays and use the allowed values of $(\delta_{+}-\delta_{-})$
from $B_s \to \phi \phi$ as an input for $B_d \to \phi K^{*0}$, we see
that $A_T^{(2)}$ is predicted to be very small in this decay (see
Fig.~\ref{fig:TPphikstAT2}).

As noted above, $\btos$ decay amplitudes can be written in terms of
$V_{tb}^*V_{ts}$ and $V_{ub}^*V_{us}$, and in pure-penguin $\btos$
decays, the $V_{ub}^*V_{us}$ term is negligible. However, this
approximation is not valid for $\btos$ transitions in which there is a
tree contribution, such as $B \to \rho K^*$. In such decays, since
$V_{ub}^*V_{us}$ has a large weak phase, the $A_T^{(i)}$'s are no
longer purely fake TP's. Thus, in order to estimate the TP's, one also
has to compute the $A_T^{(i)}$'s for the charge-conjugate decays. Now,
while it is still true that the fake $A_T^{(2)}$ TP vanishes in the
heavy-quark limit, calculating corrections to this due to the finite
$b$-quark mass is much more complicated. Because there is more than
one amplitude, the TP's depend on additional variables (magnitudes of
diagrams, weak and strong phases), and there are not enough
experimental observations to constrain the parameters. For this reason
we cannot provide a reliable estimate of the fake $A_T^{(2)}$ in this
case.

Pure-penguin $\btod$ decays are similar in this respect. The penguin
diagram proportional to $V_{ub}^*V_{ud}$ is not negligible, so there
are two amplitudes contributing to the decay. As such, the TP's depend
on more parameters than in pure-penguin $\btos$ decays, and so the
corrections to the heavy-quark limit result cannot be estimated
reliably.

\begin{figure}
\includegraphics[width=0.49\linewidth]{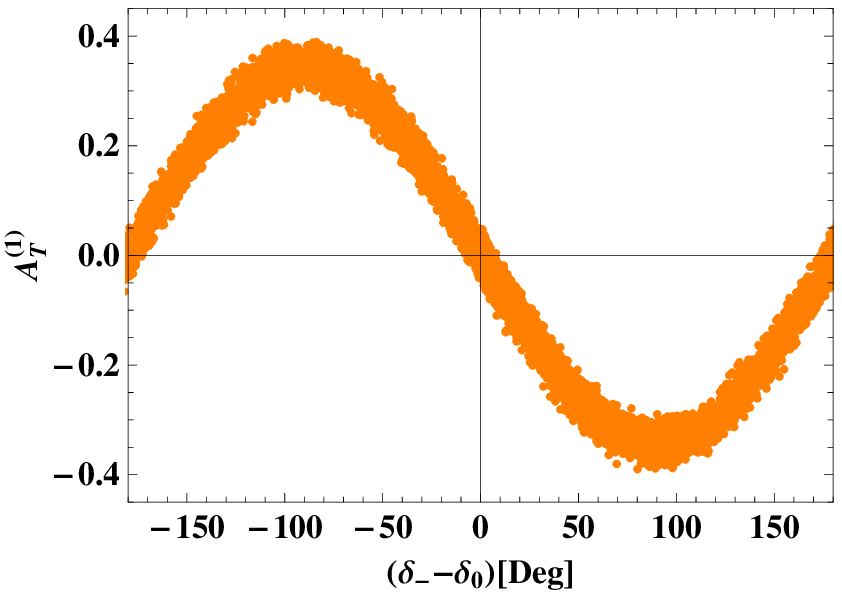}
\includegraphics[width=0.49\linewidth]{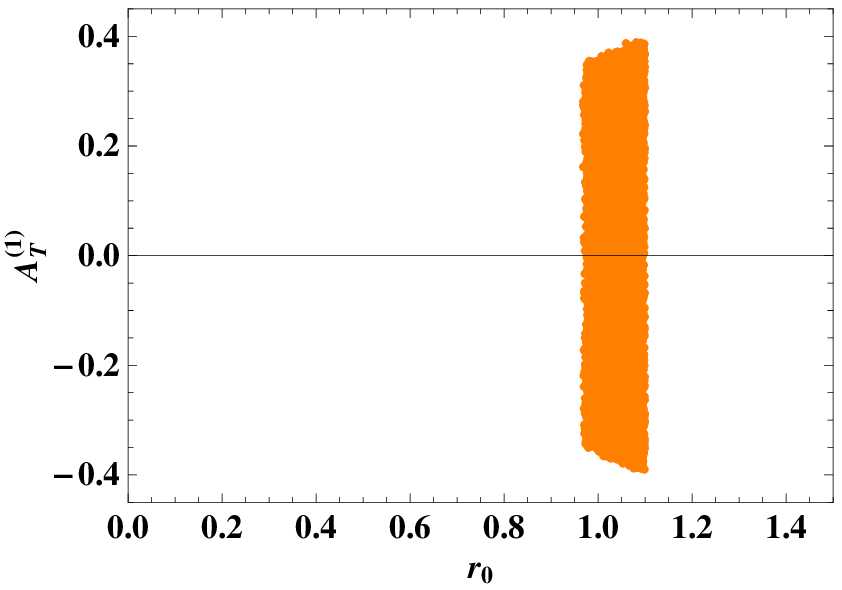}
\caption{The left (right) panel of the figure shows $A_T^{(1)}$ for
  the decay $B_d \to \phi K^{*0}$ as a function of
  $(\delta_+-\delta_-)$ ($r_T$).
  \label{fig:TPphikstAT1}}
  \end{figure}

Finally, for pure-penguin $\btos$ decays, we can estimate the
$A_T^{(1)}$ fake TP. Defining $r_0 \equiv |A_0/A_-|$, $A_T^{(1)}$ is
given by
\bea
\label{TPmeasure2}
A_T^{(1)} &\equiv&  \frac{1}{\sqrt{2}} r_0 f_L [ r_T \sin{(\delta_+-\delta_0)}-\sin{(\delta_--\delta_0)}] ~.
\eea 
$r_0$ can be fixed from
\bea
r^2_0 (1+r^2_T) &=& \frac{f_T}{f_L} ~.
\label{r0}
\eea

For $\BKstar$, we vary $r_T$ in the range (0.05, 0.15), all
phases $\delta_\lambda$ ($\lambda = 0,\pm$) in the range (0, $2\pi$),
and all polarization fractions within $\pm 1 \sigma$. This gives $r_0$
in the range (0.95, 1.1). $r_T$ and $(\delta_+-\delta_-)$ are further
constrained by Eq.~(\ref{AT2eqn}). The result for $A_T^{(1)}$ is shown
in Fig.~\ref{fig:TPphikstAT1}: $|A_T^{(1)}| \le 40\%$ is predicted.

This prediction can be compared with the experimental result, which is
deduced from other measurements as before. We find
\bea
A_T^{(1)} &=& -0.23 \pm 0.03 ~,
\eea
in agreement with the SM.

\begin{figure}
\includegraphics[width=0.49\linewidth]{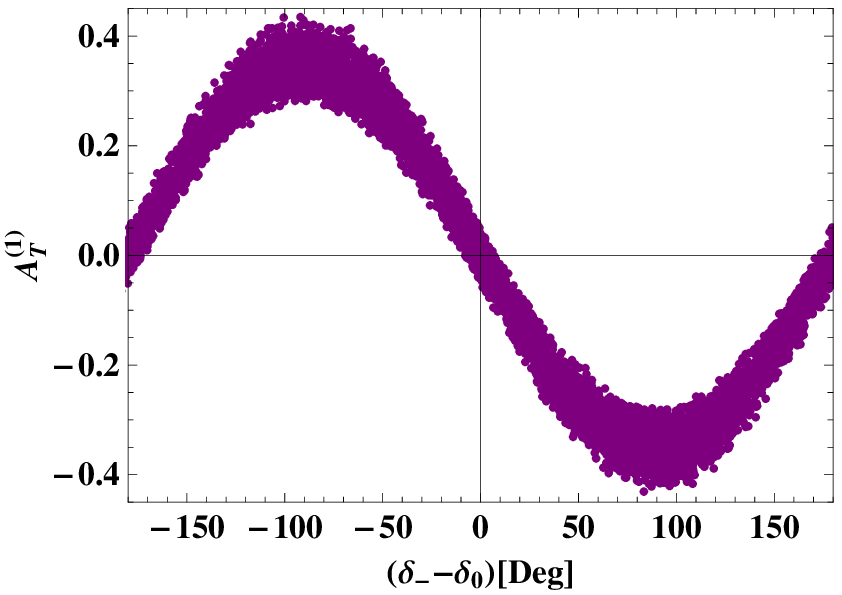}
\includegraphics[width=0.49\linewidth]{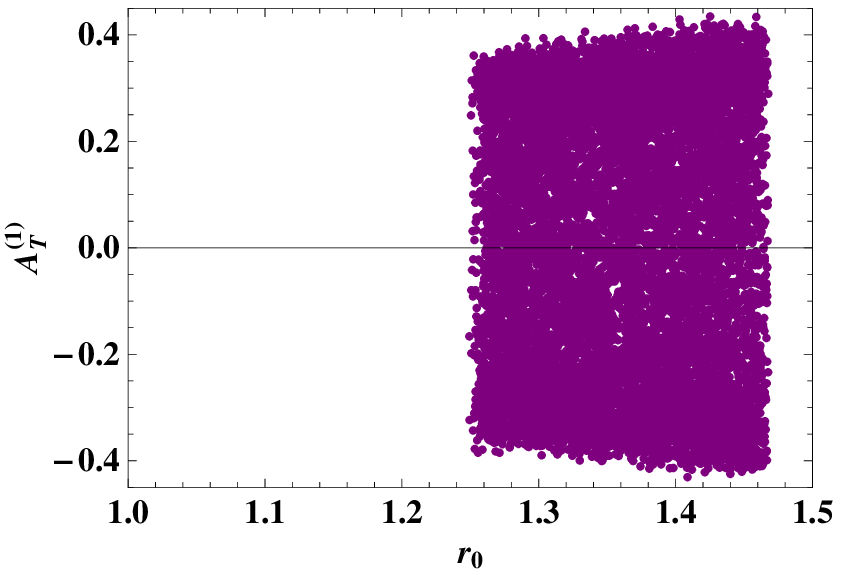}
\caption{The left (right) panel of the figure shows $A_T^{(1)}$ for
  the decay $B_s \to \phi \phi$ as a function of
  $(\delta_+-\delta_-)$ ($r_T$).
  \label{fig:TPphiphiAT1}}
\end{figure}

For $B_s \to \phi \phi$, we use the same procedure as above. $r_0$ is
found to lie in the range (1.25, 1.47). The prediction for $A_T^{(1)}$
is shown in Fig.~\ref{fig:TPphiphiAT1}. We find $|A_T^{(1)}| \le
40\%$.

In summary: the angular distribution of $B\to V_1 V_2$ ($V_{1,2}$ are
vector mesons) contains triple products (TP's), odd under parity and
time reversal. There are two TP's, denoted $A_T^{(1)} \sim {\rm
  Im}(A_\perp A_0^*)$ and $A_T^{(2)} \sim {\rm Im}(A_\perp A_\|^*)$,
where the $A_i$ ($i=0,\|,\perp$) are the polarization
amplitudes. There are TP's in ${\bar B}$ decays as well: ${\bar
  A}_T^{(i)}$ ($i=1,2$). They are equal to $-A_T^{(i)}$, with the weak
phases having the opposite sign.  There are two categories of TP's:
(i) real TP's: $\frac12[A_T^{(i)} + {\bar A}_T^{(i)}]$, and (ii) fake
TP's: $\frac12[A_T^{(i)} - {\bar A}_T^{(i)}]$.  Real TP's are
CP-violating, and are nonzero only if the decay has two contributing
amplitudes with a relative weak phase. Fake TP's are CP-conserving,
and can be generated by strong phases alone. For fake TP's, it is
necessary to distinguish $B$ and ${\bar B}$, so that tagging is
needed, possibly by using self-tagging decays.

In the heavy-quark limit, the standard model (SM) predicts that $A_\|
= -A_\perp$, so that $A_T^{(2)} = 0$. We have computed the finite-mass
corrections to this for the pure-penguin $\btos$ decays $\BKstar$ and
$\bs\to\phi\phi$.  These are especially interesting because, to a good
approximation, they have only one weak amplitude. As a consequence,
all CP-violating effects essentially vanish. In particular, these
decays have only fake TP's. For $\BKstar$, we find that the SM
predicts $|A_T^{(2)}| \le 9\%$, consistent with the measured value of
$0.002\pm 0.049$.

There is a further consequence of this measurement. In $\BKstar$, it
is expected that $\fT \ll \fL$, where $\fT$ and $\fL$ are the
fractions of transverse and longitudinal decays,
respectively. However, $\fTfL \simeq 1$ is found. Explanations of this
result have been given within the SM and with new physics
(NP). Interestingly, the NP scenarios often predict large values for
$|A_T^{(2)}|$, and are thus ruled out, or at least strongly
constrained, by the current measurement of $A_T^{(2)}$. Thus, the
measurement of the fake $A_T^{(2)}$ TP allows us to partially
differentiate the SM from the NP explanations of $\fTfL$.

We find that the SM predicts $|A_T^{(2)}| \le 10\%$ for
$\bs\to\phi\phi$. We have also estimated $A_T^{(1)}$ within the SM,
with the result that $|A_T^{(1)}| \le 40\%$ for both decays. For
$\BKstar$, this is consistent with the measured value of $-0.23 \pm
0.03$.

\bigskip
\noindent
{\bf Acknowledgments}: We are extremely grateful to Mirco Dorigo,
Marco Rescigno and Anna Maria Zanetti for asking the questions which
led to this study.  Thanks also to Marco Rescigno for pointing out an
error in an earlier version of this paper. This work was financially
supported by the US-Egypt Joint Board on Scientic and Technological
Co-operation award (Project ID: 1855) administered by the US
Department of Agricultur (AD, MD), and by NSERC of Canada (DL).


\end{document}